\documentclass[fleqn,10pt]{wlscirep}
\usepackage{cite}
\usepackage{graphics}
\usepackage{amsmath}
\usepackage{amssymb}
\usepackage{graphicx}
\usepackage{dcolumn}
\usepackage{epstopdf}
\usepackage{hyperref}

\title{Acceleration and Quantitation of Localized Correlated Spectroscopy using Deep Learning: A Pilot Simulation Study}

\author[1]{Zohaib Iqbal}
\author[1]{Dan Nguyen}
\author[2]{M. Albert Thomas}
\author[1,*]{Steve Jiang}
\affil[1]{Medical Artificial Intelligence and Automation Laboratory, Department of Radiation Oncology, University of Texas Southwestern Medical Center, Dallas, TX, USA 75390 }
\affil[2]{Department of Radiological Sciences, University of California Los Angles, Los Angeles, CA, USA}

\affil[*]{Corresponding author contact: steve.jiang@utsouthwestern.edu }

\begin{abstract}
Nuclear magnetic resonance spectroscopy (MRS) allows for the determination of atomic structures and concentrations of different chemicals in a biochemical sample of interest. MRS is used \textit{in vivo} clinically to aid in the diagnosis of several pathologies that affect metabolic pathways in the body. Typically, this experiment produces a one dimensional (1D) $^1$H spectrum containing several peaks that are well associated with biochemicals, or metabolites. However, since many of these peaks overlap, distinguishing chemicals with similar atomic structures becomes much more challenging. One technique capable of overcoming this issue is the localized correlated spectroscopy (L-COSY) experiment, which acquires a second spectral dimension and spreads overlapping signal across this second dimension. Unfortunately, the acquisition of a two dimensional (2D) spectroscopy experiment is extremely time consuming. Furthermore, quantitation of a 2D spectrum is more complex. Recently, artificial intelligence has emerged in the field of medicine as a powerful force capable of diagnosing disease, aiding in treatment, and even predicting treatment outcome. In this study, we utilize deep learning to: 1) accelerate the L-COSY experiment and 2) quantify L-COSY spectra. All training and testing samples were produced using simulated metabolite spectra for chemicals found in the human body. We demonstrate that our deep learning model greatly outperforms compressed sensing based reconstruction of L-COSY spectra at higher acceleration factors. Specifically, at four-fold acceleration, our method has less than 5\% normalized mean squared error, whereas compressed sensing yields 20\% normalized mean squared error. We also show that at low SNR (25\% noise compared to maximum signal), our deep learning model has less than 8\% normalized mean squared error for quantitation of L-COSY spectra. These pilot simulation results appear promising and may help improve the efficiency and accuracy of L-COSY experiments in the future. 
\end{abstract}

\begin{document}

\flushbottom
\maketitle
%
%
\thispagestyle{empty}

\noindent

\section*{Introduction}
Magnetic resonance imaging (MRI) is a popular imaging modality capable of providing valuable anatomical and functional information \textit{in vivo}. By utilizing a strong magnetic field and radio-frequency (RF) waves, MRI successfully images hydrogen atoms in their local chemical environment, allowing for useful soft tissue contrast. One technique that allows for the metabolic investigation of different tissues is the magnetic resonance spectroscopy (MRS) method. In particular, single-voxel $^1$H MRS is capable of providing biochemical information from a volume of interest (VOI) in the human body\cite{bottomley1987spatial}. MRS provides a $^1$H spectrum rich with peaks representative of various chemicals. Furthermore, this spectrum can be quantified by using a spectral fitting algorithm\cite{provencher1993estimation,ratiney2005time,naressi2001java,vanhamme1997improved,wilson2011constrained} to yield chemical, or metabolite, concentrations. MRS, and more specifically the point resolved spectroscopy (PRESS) experiment, has been used to explore pathologies affecting the brain\cite{soares2009magnetic}, prostate\cite{costello1999citrate}, liver\cite{fischbach2008assessment}, breast\cite{bolan2005imaging}, as well as other sites, and is often used in combination with other imaging studies to discern how metabolic alterations in tissues correlate with anatomical abnormalities. 

Unfortunately, one-dimensional (1D) spectroscopy techniques such as PRESS have a disadvantage when it comes to quantifying overlapping metabolite spectral signals. Since many metabolites are found in the body at very low concentrations, separating these signals from more dominant spectral peaks becomes very challenging. For this reason, several approaches have been developed to better quantify these lower concentrated metabolites, including J-editing techniques\cite{mescher1998simultaneous,edden2012macromolecule,chan2016hermes,chan2017echo} and two-dimensional (2D) spectral acquisitions\cite{aue1976two,ryner1995localized,kreis1996spatially,thomas2001localized,dreher1999detection}. In particular, 2D MRS offers the advantage of quantifying all metabolite signals in a single scan at the expense of increasing acquisition time. A typical 2D MRS experiment includes a time increment, $t_1$, in the pulse sequence to acquire data from the indirect temporal dimension. Combined with the acquisition of the direct temporal dimension, $t_2$, a 2D spectrum, $S$($F_2$,$F_1$), can be acquired by Fourier transforming the 2D temporal data, $s$($t_2$,$t_1$). 

One popular 2D MRS technique is the localized correlated spectroscopy (L-COSY) experiment\cite{thomas2001localized}. This experiment acquires data by using a 90$^\circ$-180$^\circ$-$t_1$-90$^\circ$-$t_2$ sequence and yields several cross-peaks which can be used to identify and quantify overlapping resonances. However, there are two main limitations of the L-COSY technique. First, due to the $t_1$ increment necessary to obtain the indirect dimension, the L-COSY scan time is very long. Second, because of the nature of an additional dimension, spectral fitting becomes more complex and therefore less ideal quantitation techniques such as peak integrals are often used. Several methods have been proposed to overcome these two challenges to improve L-COSY, including non-uniform sampling with reconstruction\cite{schmieder1993application} and 2D spectral fitting using prior-knowledge\cite{schulte2006profit,martel2015localized}.

Recently, deep learning and artificial intelligence have become more prominent in the medical field and radiology\cite{lecun1989backpropagation,lecun2015deep,goodfellow2016deep,ronneberger2015u}. These methods are often used for segmenting medical images, aiding with diagnosis, and verifying image quality. One popular deep learning architecture is the UNet\cite{ronneberger2015u}, which is a fully convolutional network\cite{long2015fully} capable of image-to-image domain mapping. While UNet is often used for segmentation purposes, our group has recently demonstrated that a novel UNet architecture, the densely connected U-Net (D-UNet)\cite{huang2017densely,iqbal2018super,nguyen2018three}, is capable of reconstructing super-resolution spectroscopic images. In this study, we demonstrate that the D-UNet architecture can be used to: 1) reconstruct non-uniformly sampled (NUS) L-COSY acquisitions and 2) quantify fully sampled L-COSY spectra accurately. The D-UNet models were trained and evaluated using simulated L-COSY data. The first type of D-UNet model was trained to reconstruct NUS L-COSY. This reconstruction method was quantitatively compared to compressed sensing ($\ell_1$-norm) reconstruction\cite{lustig2007sparse}. The second type of D-UNet model was trained to quantify seventeen metabolites from a simulated fully sampled L-COSY spectrum. All reconstruction results were compared to the actual simulations to evaluate the errors of the reconstructions both qualitatively and quantitatively.  

\section*{Methods} 
\begin{figure}
	\center
	\hspace*{-0.5cm}
	\includegraphics[scale=0.22]{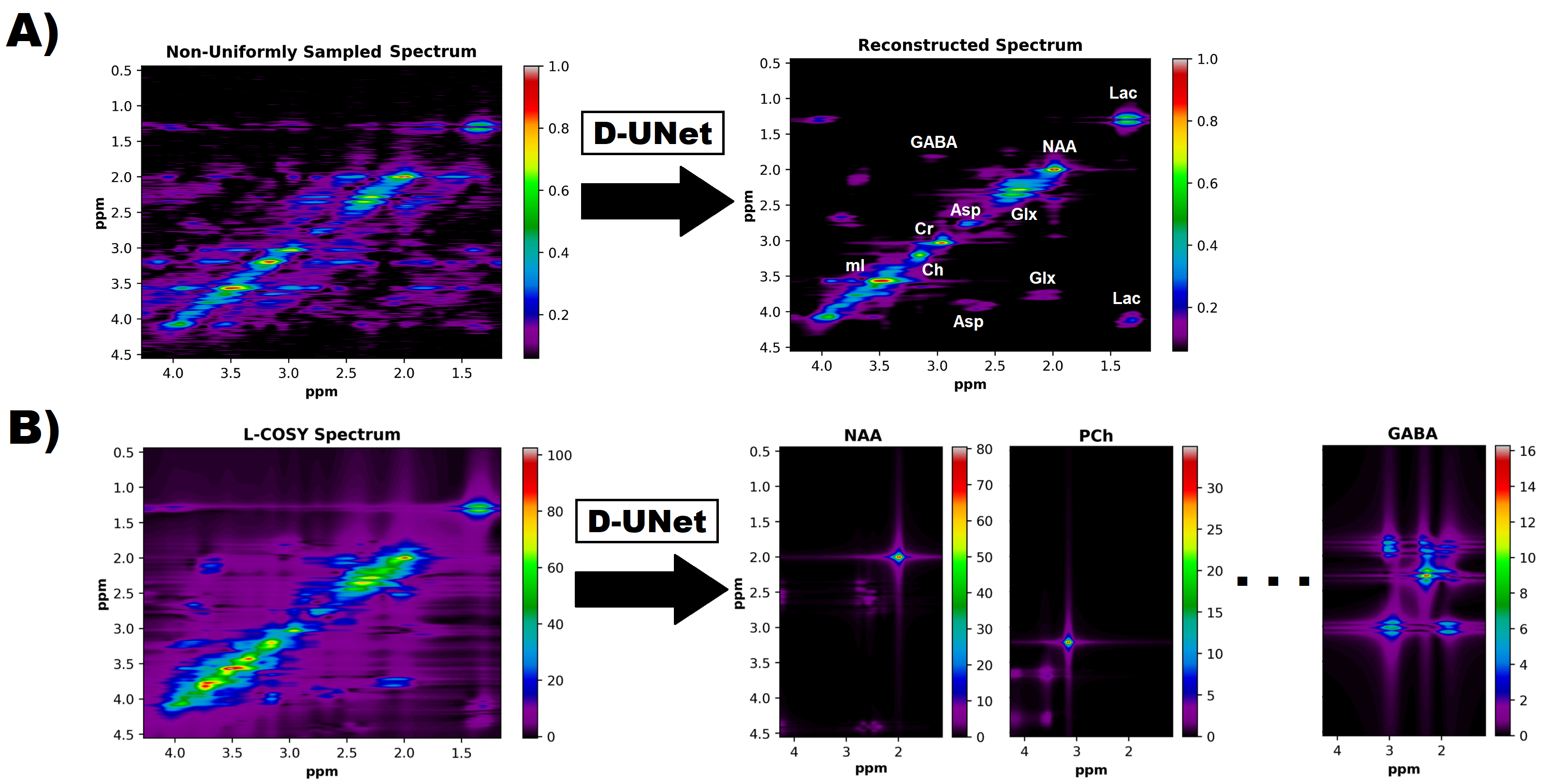}
	\caption{\small Two proposed implementations for the D-UNet architecture are shown. A) A non-uniformly sampled L-COSY experiment is reconstructed into the fully sampled spectrum. While under-sampling can be performed in both the $t_2$ and $t_1$ dimensions, this study analyzes the reconstruction of L-COSY spectra acquired using non-uniform sampling along only the $t_1$ dimension. B) Several metabolite spectra are identified from a fully sampled L-COSY spectrum using a D-UNet model. The intensities of the metabolite spectra directly correlate to concentration values, and therefore this study also investigates the potential application of deep learning to quantify L-COSY spectra. In total, 17 metabolites were quantified in this simulation study.}
	\label{fig:flow}
\end{figure}

As shown in Figure \ref{fig:flow}, the goal of this study was to perform two distinct tasks using the D-UNet architecture: 1) reconstruct NUS L-COSY spectra and 2) quantify L-COSY spectra. While each task used different data for training the models and testing the results, the initial simulation process to synthesize L-COSY spectra was identical for both applications. 

\subsection*{Simulation}
GAMMA simulation\cite{smith1994computer} was used to simulate seventeen different metabolites found in the human brain using the 90$^\circ$-180$^\circ$-$t_1$-90$^\circ$-$t_2$ L-COSY sequence\cite{thomas2001localized}. These metabolites included aspartate (Asp), choline (Ch), creatine at 3ppm (Cr3.0), creatine at 3.9ppm (Cr3.9), $\gamma$-butyric acid (GABA), glucose (Glc), glutamine (Gln), glutamate (Glu), glutathione (GSH), lactate (Lac), myo-Inositol (mI), N-acetyl aspartate (NAA), N-acetyl-asparate-g (NAAG), phosphocholine (PCh), phosphoethanolamine (PE), taurine (Tau), and threonine (Thr). Chemical shift values for the biochemicals were found in the literature\cite{govindaraju2000proton}. The metabolites were simulated using the following experimental parameters: TE=30ms, $t_2$ points = 2048, $t_1$ points = 100, spectral bandwidth along the direct dimension ($SBW_2$) = 2000Hz, and spectral bandwidth along the indirect dimension ($SBW_1$) = 1250Hz. The magnetic field strength was chosen to be the field strength of a Siemen's 3T scanner (Erlangen, Germany). 

Then, L-COSY spectra were randomly generated by modifying the original metabolite simulations, also referred to as the basis set. Each metabolite in the basis set ($B_m$) was first line broadened in both the direct and indirect temporal dimensions using an exponential filter and a random phase was applied to the basis metabolite signal as well: 

\begin{equation} \label{eq:lb}
B_{lb,m} = B_m e^{-r_{1,m}} e^{-r_{2,m}} e^{-i\phi_r}
\end{equation} 

Above, $B_{lb,m}$ is the new line-broadened metabolite, $\phi_r$ is a random angle between 0 and 2$\pi$, $e^{-r_{1,m}}$ is an exponential filter applied to the $t_1$ domain, and $e^{-r_{2,m}}$ is an exponential filter applied to the $t_2$ domain. Each metabolite was allowed to have separate line-broadening terms. The factors $e^{-r_{2,m}}$ and $e^{-r_{1,m}}$ resulted in effective line-broadenings of 5-25Hz and 0-15Hz, respectively, and were implemented in this fashion to mimic the range of common $T_2$ values \textit{in vivo}.    

Next, the individual metabolites were combined linearly using random concentration values to produce an initial L-COSY spectrum, $s_{init}$:

\begin{equation} \label{eq:add}
s_{init} = \sum_{m} r_{3,m} B_{lb,m}
\end{equation}   

In equation \ref{eq:add}, $r_{3,m}$ is a random concentration value between 0 and 10, and is representative of the concentration value in mmol. The final L-COSY spectrum, $s_f$, was created by adding noise to $s_{init}$. The noise level could vary drastically from 0\% to 25\% of the maximum metabolite signal.

\subsection*{Non-uniform Sampling and Reconstruction}     
\begin{figure}
	\center
	\hspace*{-1cm}
	\includegraphics[scale=0.38]{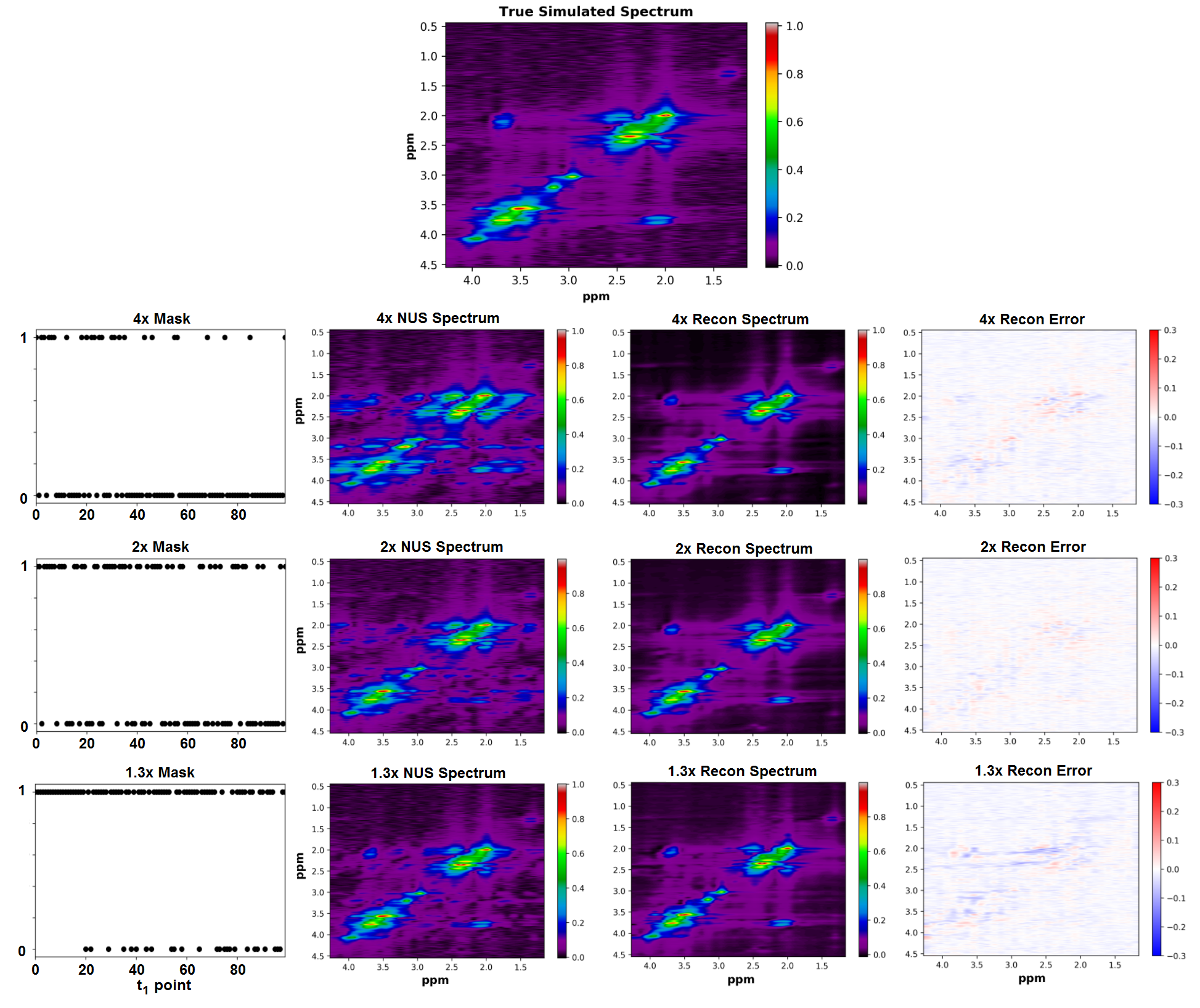}
	\caption{\small A ground truth simulated L-COSY spectrum is shown (top). Sampling schemes were applied to the simulated spectrum using the sampling masks shown in the 1$^{st}$ column. These masks sampled 25, 50, and 75 $t_1$ points out of a total 100 $t_1$ points to yield 4x, 2x, and 1.3x acceleration factors, respectively. The 2$^{nd}$ column shows the under-sampled spectra in the ($F_2$,$F_1$) domain and the 3$^{rd}$ column shows the spectra reconstructed using a D-UNet model. Errors for each reconstruction are displayed as difference maps in the final column.}
	\label{fig:NUSrecon}
\end{figure}

Non-uniform sampling was performed on the final $s_f$ matrix along the $t_1$ dimension utilizing an exponential probability density function\cite{macura1983improved,wilson2015accelerated,iqbal20163d}. This NUS scheme emphasized sampling earlier $t_1$ points more due to the fact that these points have less $T_2$ decay (more signal). The last $t_1$ point was sampled for all of the NUS schemes. The three sampling masks used in this study are displayed in Figure \ref{fig:NUSrecon}. A $t_1$ point was sampled if the value in the mask was 1, and it was not sampled if the value in the mask was 0. The number of points sampled for each mask were 75, 50, and 25 resulting in a scan acceleration factor of 1.3x, 2x, and 4x, respectively. 

Aside from the D-UNet reconstruction of NUS data described below, data were also reconstructed using compressed sensing reconstruction\cite{lustig2007sparse}. The $\ell_1$-norm minimization reconstruction was performed by solving the following optimization problem:

\begin{equation} \label{eq:ell}
\begin{aligned}
& \underset{u}{\text{minimize}}
& & ||u||_{_1} \\
& \text{subject to}
& & ||MFu - f||^2_2 \leq \sigma^2
\end{aligned} 
\end{equation}

Equation \ref{eq:ell} is the general formulation for compressed sensing reconstruction. $u$ is the reconstructed data in the ($F_2$,$F_1$) spectral domain, $M$ is the sampling mask along the $t_1$ domain, $F$ is the 2D Fourier transformation, $f$ is the NUS data in the ($t_2$,$t_1$) temporal domain, and $\sigma^2$ is the estimate of the noise variance. The noise variance was estimated from a noisy region of the spectrum, as previously described\cite{wilson2015accelerated,wilson2015correlated,burns2014group,burns2014non}.  

\begin{figure}
	\center
	\hspace*{1.0cm}
	\includegraphics[scale=0.25]{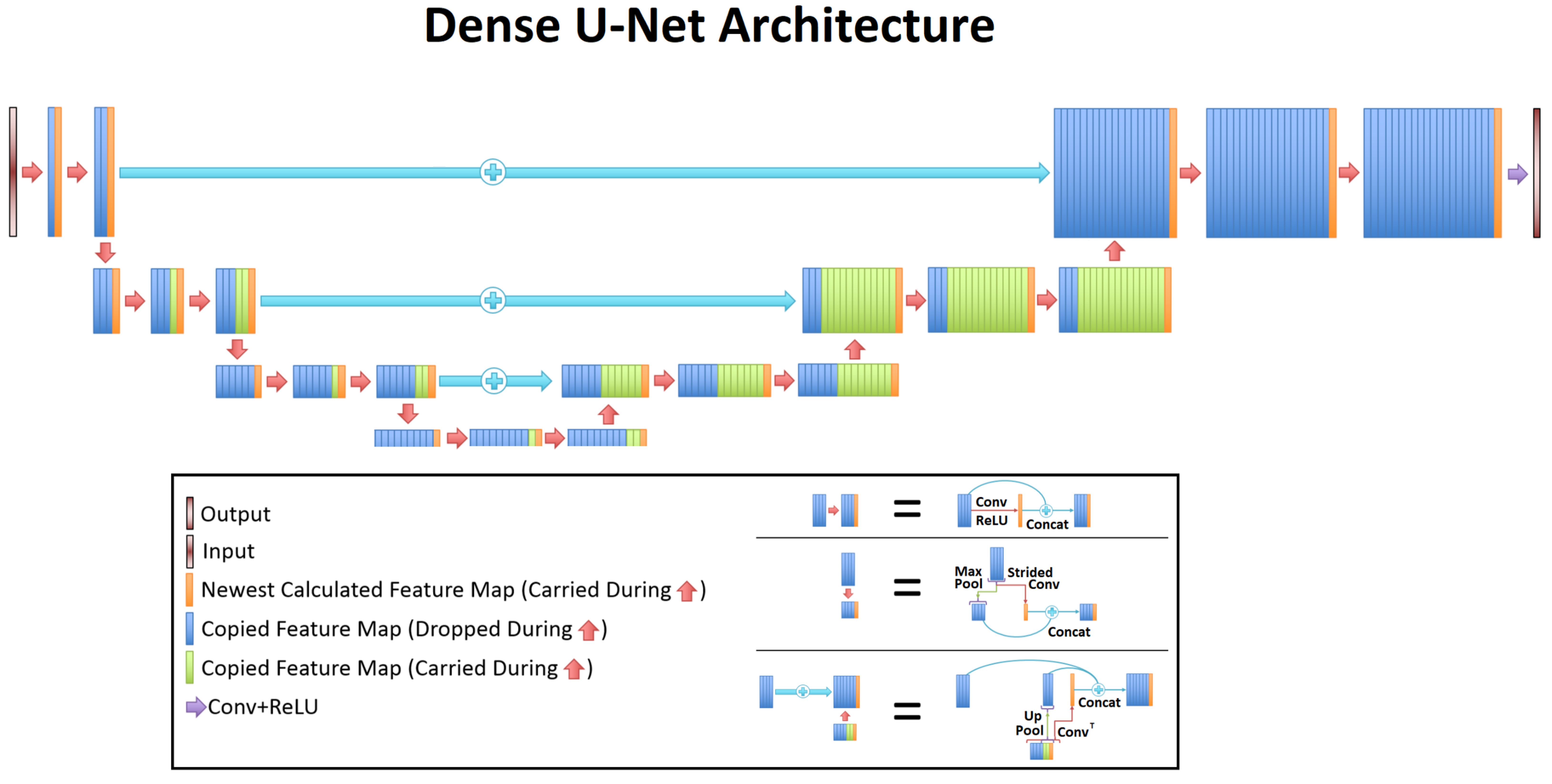}
	\caption{\small The densely connected U-Net architecture from a previous publication\cite{iqbal2018super} is displayed. The densely connected flavor of this model allows for important features to be carried over throughout the entire training process.}
	\label{fig:model}
\end{figure}

\subsection*{Reconstructing NUS L-COSY with D-UNet}
The densely connected UNet architecture utilized in this study was very similar to a previously reported model\cite{iqbal2018super}, and the general architecture can be seen in figure \ref{fig:model}. This model utilized the generic UNet architecture, which operates by learning important global and local features using a variety of convolutional layers. The first half of the UNet continuously uses convolutional and max pooling layers, and these layers help reduce the input matrix size. By reducing the size, the network learns the primary global features of the input images. The second half of the UNet uses deconvolutional and up-pooling layers, which restore the matrix size. This process helps learn local features that are vital to restoring the images on a finer scale. The architecture also leveraged densely connected convolutional layers, which aid in carrying important features throughout the learning process. All convolutional layers used a kernel size = 3 x 3, stride = 1, and a rectified linear unit (ReLU) activation function\cite{krizhevsky2012imagenet}.

The D-UNet model used for reconstructing the NUS L-COSY data was designed to take an NUS L-COSY spectrum as input and produce a reconstructed L-COSY spectrum as output. The NUS L-COSY data was produced by multiplying $s_f$ by the sampling mask in the ($F_2$,$t_1$) domain and then transforming this matrix back into the ($F_2$,$F_1$) domain. The output was simply the $s_f$ matrix without noise in the ($F_2$,$F_1$) domain. Both the input and output matrix sizes were 512 x 32, and corresponded to spectral ranges of 0.5-4.5ppm in the direct spectral dimension ($F_2$), and 1.2-4.3ppm in the indirect spectral dimension ($F_1$). Additionally, the inputs and outputs were inserted as three different channels into the network with each channel representing the real, imaginary, and magnitude information of the spectrum. Finally, all inputs and outputs were normalized to be in between values of 0 and 1, and were normalized based on the maximum value of the magnitude images. The loss function was the mean squared error (MSE) between the reconstructed L-COSY (Recon) and the actual simulated L-COSY (Actual), which was defined as:

\begin{equation} \label{eq:mse}
MSE = \sum_{F_2}\sum_{F_1}\frac{(Recon - Actual)^2}{512*32}
\end{equation}

The Adam optimizer\cite{kingma2014adam} was used with a learning rate set to 1e$^{-3}$. Three D-UNet models with identical architecture were trained to reconstruct spectra sampled using the masks shown in figure \ref{fig:NUSrecon}. A total of 40,000 simulated NUS L-COSY spectra were simulated for each sampling scheme, and 100 spectra were used to evaluate the results as an independent test set. The batch size for the training was 10 samples per batch.

\subsection*{Quantitation of L-COSY with D-UNet}
The quantitation of fully sampled L-COSY data was performed in a similar manner to the method described above. The input to the quantitation D-UNet was the L-COSY spectrum as a 512 x 32 matrix with three channels representing the magnitude, real, and imaginary components of the spectrum. The input was scaled from 0 to 100 based on the maximum of the magnitude spectrum. The output of the network was a 512 x 32 matrix representative of each metabolite basis set. Therefore, since 17 metabolites were quantified, the output had 17 channels representing the magnitude spectrum for each metabolite. All other training parameters were identical to those described above. A total of 21,000 simulated L-COSY spectra were used for training, and 100 spectra were used for testing the results independently. 

\subsection*{Evaluation}
All of the results were compared to the actual simulated spectra by utilizing the MSE metric from equation \ref{eq:mse}. For the non-uniformly sampled spectral reconstruction, the MSE was calculated over all 100 test spectra and compared to the MSE of the $\ell_1$-norm reconstruction from equation \ref{eq:ell} for all acceleration factors. Normalized MSE was also used, and errors were normalized based on the maximum signal intensity of the spectrum. In addition to MSE, the quantitation with D-UNet also investigated the effect of noise on the quantitative results. Specifically, ten different noise levels were evaluated on the same 100 spectra to determine how the signal-to-noise ratio (SNR) affects the model results and overall stability. These noise levels ranged from 0\% to 25\% of the maximum signal intensity.   

\section*{Results}
\subsection*{NUS L-COSY reconstruction}
\begin{figure}
	\center
	\hspace*{-0.5cm}
	\includegraphics[scale=0.5]{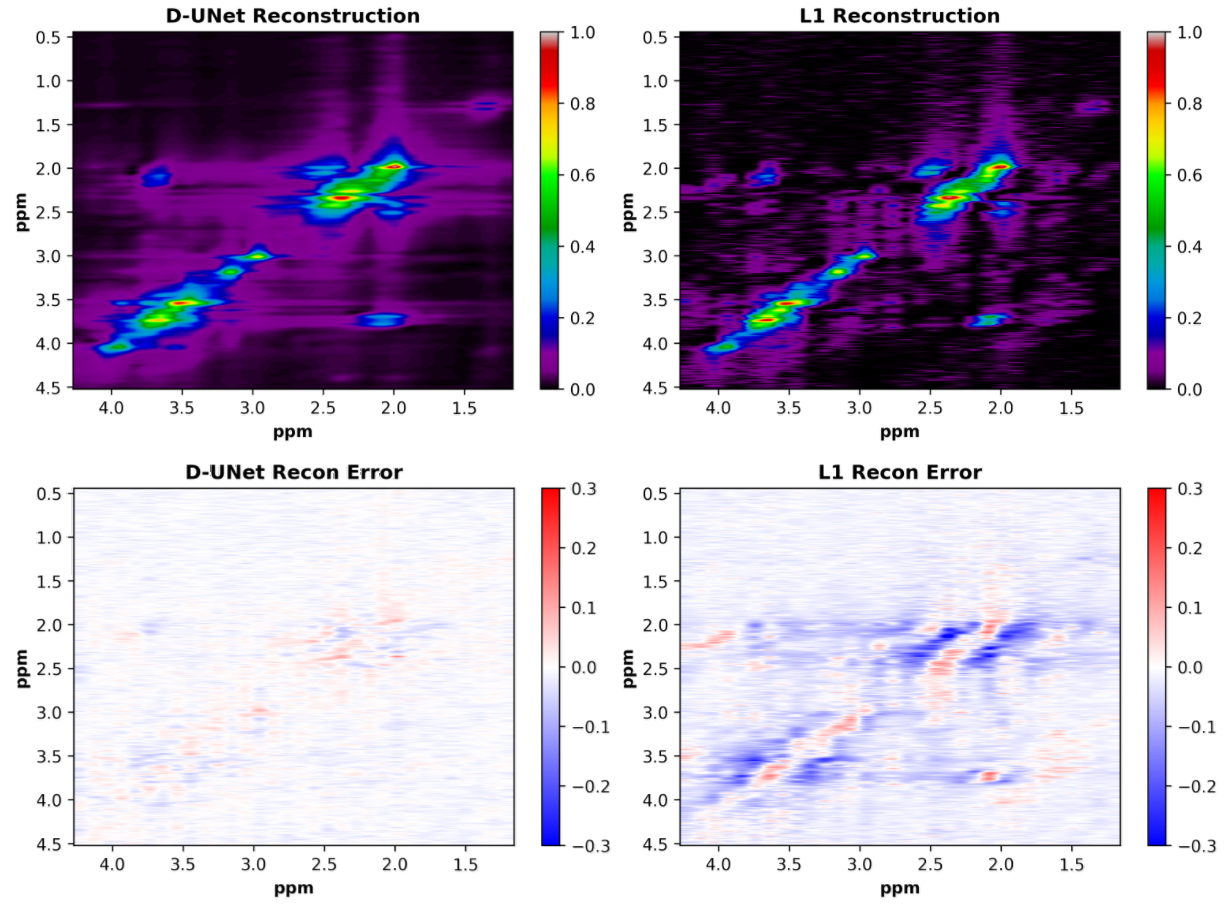}
	\caption{\small A qualitative comparison between the D-UNet and the compressed sensing ($\ell_1$-norm) reconstructions is shown. The fully sampled L-COSY spectrum displayed in figure \ref{fig:NUSrecon} was sampled using 25 $t_1$ points (4x acceleration). The spectrum was then reconstructed using the trained deep learning model and optimization described in equation \ref{eq:ell}. Errors between the two reconstructions are displayed as differences between the actual spectrum and the reconstructions.}
	\label{fig:NUScomp}
\end{figure}

The NUS L-COSY spectra reconstructed using the D-UNet architecture can be seen in figure \ref{fig:NUSrecon}. The non-uniform sampling produces several $F_1$ ridging artifacts present in the spectral domain, which are ultimately removed by using the D-UNet models. For training, the MSE loss function achieved a loss of approximately 3e$^{-5}$ for each of the models. The errors as the difference between the Actual and Recon spectra are also shown for each acceleration factor.

\begin{table}[!h]
	\resizebox{2\textwidth}{!}{\begin{minipage}{\textwidth}
\begin{tabular}{ccc}
	\toprule
	Acceleration & D-UNet & $\ell_1$ \\ \midrule
	1.3x & 0.0388 & \textbf{0.0192}\\
	2x & \textbf{0.0170} & 0.0681   \\
	4x & \textbf{0.0327}  & 0.208    \\ \bottomrule
\end{tabular}
\end{minipage}}
\caption{\small Total mean squared error (MSE) over 100 testing spectra for each acceleration factor. A D-UNet model was trained to learn reconstruction for each sampling factor, and the results were compared to $\ell_1$-norm reconstruction as described in equation \ref{eq:ell}. Since the maximum signal is 1 for all spectra analyzed, the normalized MSE as a percentage for the D-UNet is 3.88\%, 1.70\%, and 3.27\% for acceleration factors of 1.3x, 2x, and 4x, respectively. Similarly, the MSE as a percentage for the $\ell_1$-norm reconstruction is 1.92\%, 6.81\%, and 20.8\% for acceleration factors of 1.3x, 2x, and 4x, respectively.}
\label{table:MSE}
\end{table}

A qualitative comparison between the D-UNet reconstruction and $\ell_1$-norm minimization methods are shown in figure \ref{fig:NUScomp} for the 4x reconstructions. While the D-UNet reconstruction displays minimal errors surrounding the major peaks, the compressed sensing results show large errors. Due to the iterative reconstruction, several false cross-peaks also appear in the $\ell_1$-norm reconstructed spectra, which are not present in the D-UNet reconstruction. Also, a quantitative comparison between the two reconstruction methods is provided in table \ref{table:MSE}. At lower acceleration factors where more points are sampled, $\ell_1$-norm minimization performs better than the D-UNet reconstruction. However, at higher acceleration factors where less points are sampled, the D-UNet mean error remains under 5\%, whereas the $\ell_1$-norm minimization reconstruction error is larger than 20\%. Once again, these values were calculated over 100 testing L-COSY data that were simulated independently of the training set.

\subsection*{L-COSY quantitation}

\begin{figure}
	\center
	\hspace*{-0.5cm}
	\includegraphics[scale=0.4]{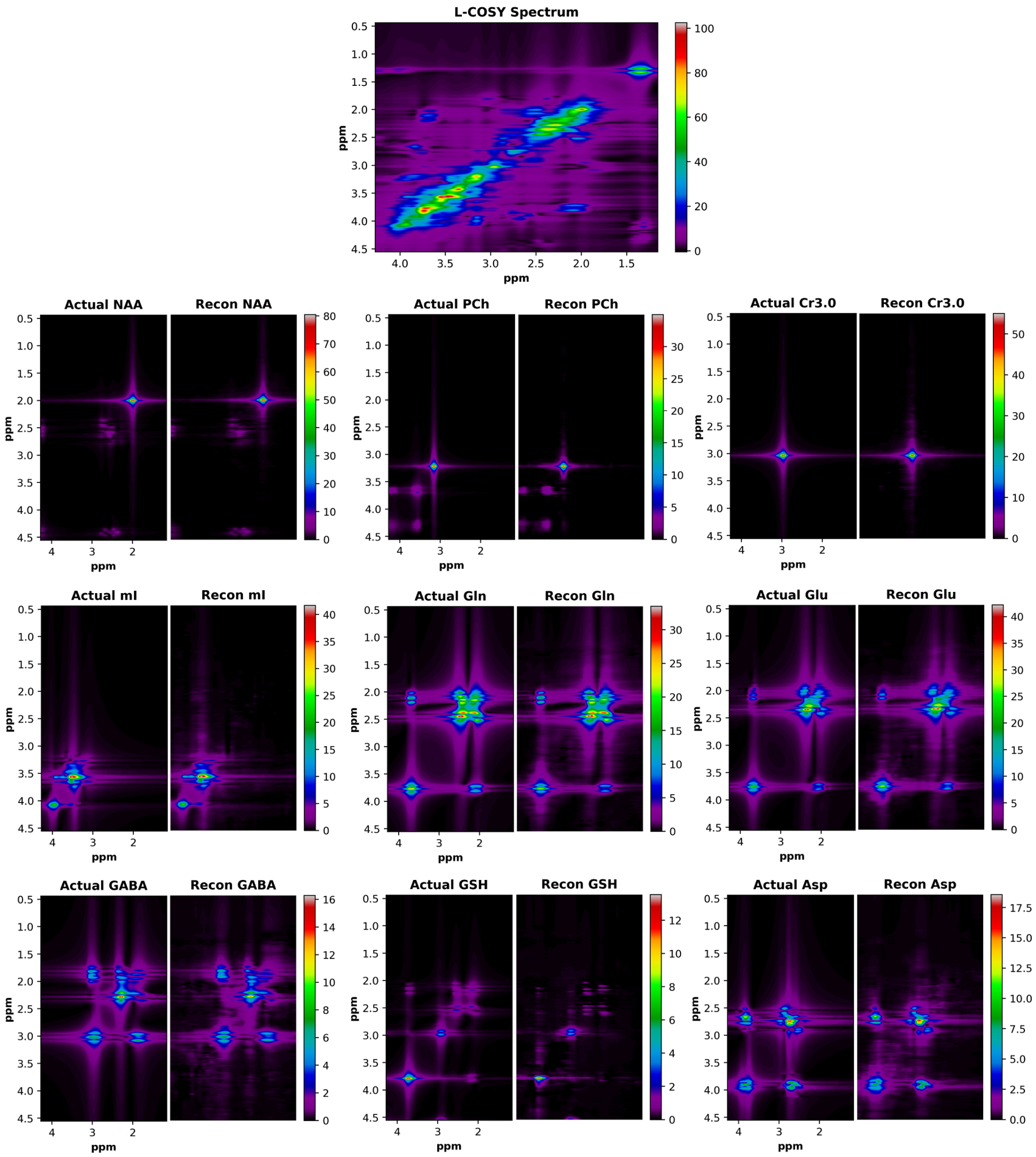}
	\caption{\small The results for the quantitation D-UNet are displayed for an example fully sampled L-COSY spectrum. From the input spectrum (top), the deep learning model reconstructs each metabolite's magnitude spectrum individually (Recon). For comparison, the actual simulated magnitude spectra (Actual) are plotted alongside the reconstructed spectra with the same intensity windows. While only 9 metabolites are displayed, the D-UNet model produces 17 metabolite spectra. The concentrations for these spectra are proportional to the signal intensities, as is standard for most fitting algorithms.}
	\label{fig:quant}
\end{figure}

The capabilities of the D-UNet to identify metabolites from a given L-COSY spectrum are demonstrated in figure \ref{fig:quant}. From the given L-COSY spectrum, 9 metabolite reconstructions are shown and compared alongside the simulated ground truth spectra: NAA, PCh, Cr3.0, mI, Gln, Glu, GABA, GSH, and Asp. In the example spectrum displayed, NAA was simulated at a concentration level of approximately 8 mmol. For GSH, which was simulated closer to 1 mmol, the reconstruction results still have similar intensity values to the simulated ground truth. While only 9 metabolite reconstructions are shown, it is important to note that all 17 metabolites in the basis set are reconstructed and could be visualized. 

\begin{figure}
	\center
	\hspace*{-0.5cm}
	\includegraphics[scale=0.28]{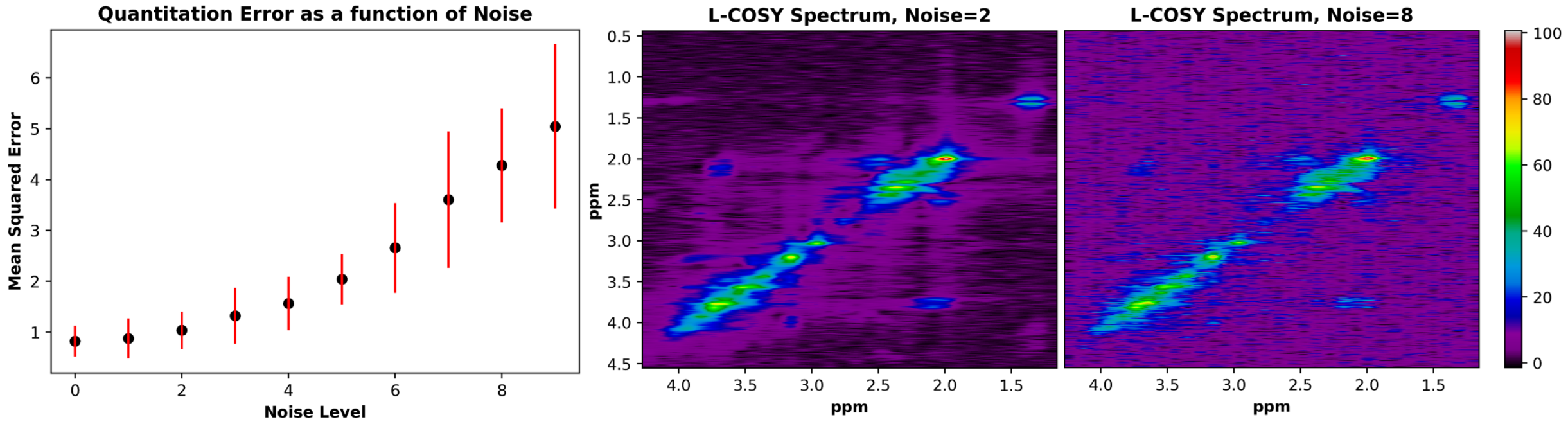}
	\caption{\small The mean squared error is displayed as a function of noise level for the quantitation D-UNet results (left). These results were produced by analyzing MSE for 100 identical spectra at 10 different noise levels ranging from 0\% - 25\% noise relative to the maximum signal intensity. Two example spectra are shown displying 5\% noise (middle) and 20\% noise (right). Qualitatively, it is clear that cross-peak signal amplitude is greatly altered due to the added noise in the noise level = 8 spectrum.}
	\label{fig:MSEnoise}
\end{figure}

Of course, SNR can play a large role on the performance of any quantitation algorithm, and therefore errors resulting from high noise were investigated. Figure \ref{fig:MSEnoise} displays the effect of noise levels on the calculated mean squared error for all 17 metabolite spectral reconstructions. As expected, degrading SNR results in larger MSE values for quantitation. In addition, an example spectrum is shown at two different noise levels: noise level 2 (5\% noise) and noise level 8 (20\% noise). It is clear that cross-peak intensities vary largely with noise, due to the fact that cross-peaks are low signal peaks for the L-COSY experiment.  

\begin{figure}
	\center
	\hspace*{-0.5cm}
	\includegraphics[scale=0.33]{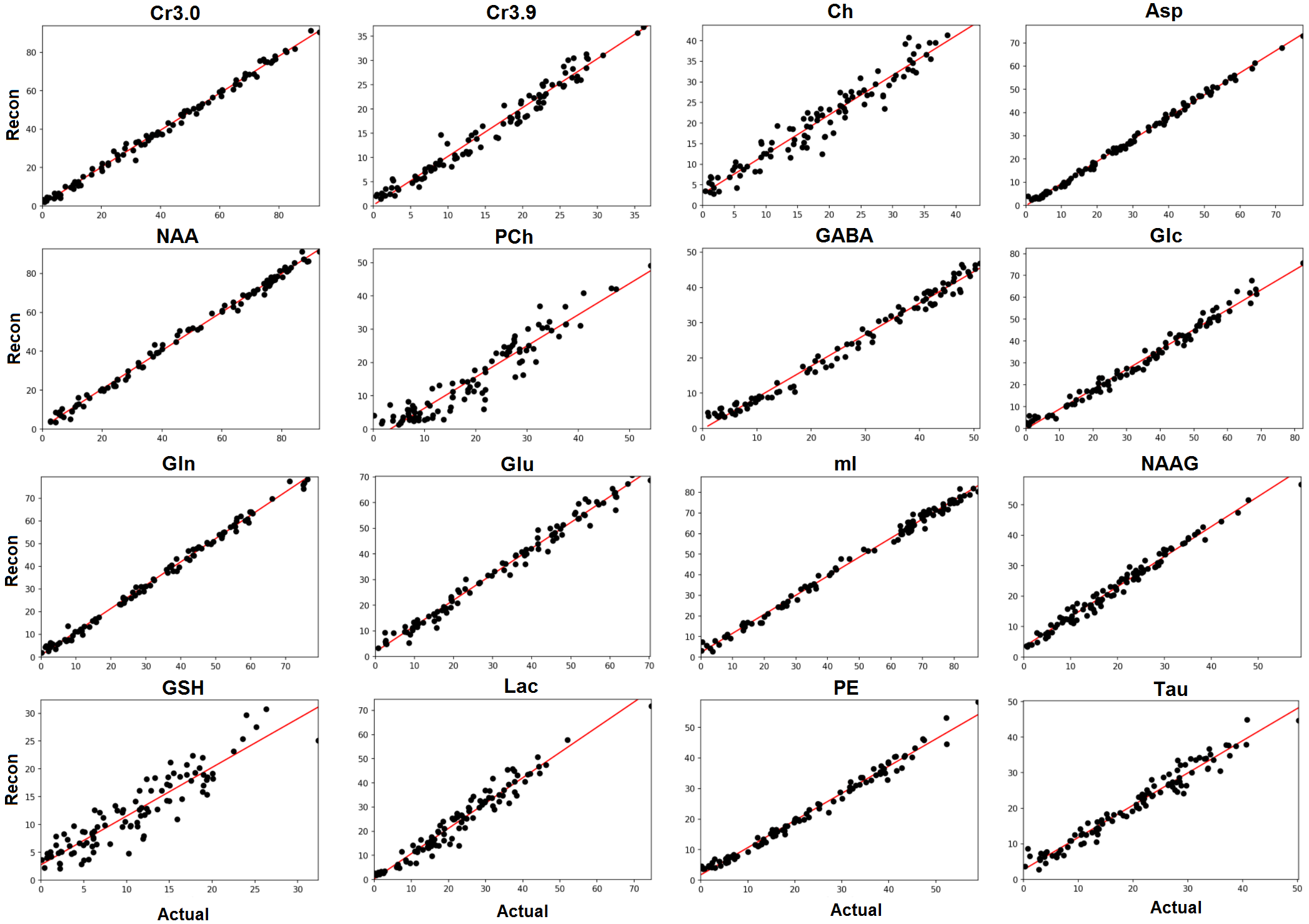}
	\caption{\small The relationship between the actual metabolite concentration on the x-axis (Actual) and the reconstructed metabolite concentration on the y-axis (Recon) is shown for 16 metabolites for 100 test spectra. The spectra contained approximately 5\% noise signal relative to the maximum signal intensity (noise level = 2). Overall, most metabolites displayed an expected linear relationship even at lower concentration values. Quantitative values for these results are tabulated in table \ref{table:fittab}.}
	\label{fig:fit}
\end{figure}

The linear relationships between the actual and predicted measurements for all 100 test spectra and 17 metabolites were also analyzed. Figure \ref{fig:fit} shows the linear relationships for 16 metabolites quantified from the test spectra with a noise level of 5\% of the maximum signal intensity. Linear fits are shown on the correlation plots between the simulated ground truth (Actual) and the reconstructed (Recon) concentrated values. In order to produce the concentration results for Recon, the maximum intensity was used from the individually reconstructed metabolite spectra from the D-UNet quantitation model. 

Finally, table \ref{table:fittab} compares the concentration values of the 17 metabolites at different SNR values. Ideally, if the quantitation was perfect, the slope would be one and the standard error would be zero. For many metabolites at noise level=2, the slope and error are close to ideal values. However, at noise level=8, slopes start to deviate largely from the ideal values and error also increases. The r$^2$ metric displayed is the coefficient of determination and is the variance of the fit. Overall, the r$^2$ values show that variance is low for the quantitative correlations at both noise levels, as demonstrated by r$^2>$0.8.

\begin{table}[!h]
	\resizebox{1.6\textwidth}{!}{\begin{minipage}{\textwidth}
\begin{tabular}{c|ccc|ccc}
	\toprule
	 & \multicolumn{3}{c}{\textbf{Noise Level = 2}} & \multicolumn{3}{c}{\textbf{Noise Level = 8}} \\ 
	 Metabolite & Slope & $r^2$ & Std. Error & Slope & $r^2$ & Std. Error\\ \midrule
	 Asp & 0.939 & 0.997 & 0.00734 & 0.683 & 0.946 & 0.0235\\
	 Ch & 0.948 & 0.974 & 0.0224 & 0.747 & 0.842 & 0.0484\\
	 Cr3.0 & 0.970 & 0.998 & 0.00667 & 0.849 & 0.949 & 0.0285\\ 
	 Cr3.9 & 1.02 & 0.974 & 0.0241 & 0.561 & 0.740 & 0.0516\\ 
	 GABA & 0.915 & 0.989 & 0.0136 & 0.591 & 0.854 & 0.0363\\
	 Glc & 0.921 & 0.992 & 0.0120 & 0.804 & 0.927 & 0.0328\\
	 Gln & 1.04 & 0.997 & 0.00846 & 0.723 & 0.913 & 0.0327\\
	 Glu & 1.01 & 0.995 & 0.0101 & 0.873 & 0.933 & 0.0342\\
	 GSH & 0.936 & 0.940 & 0.0344 & 0.696 & 0.599 & 0.0941\\
	 Lac & 1.03 & 0.974 & 0.0242 & 0.617 & 0.806 & 0.0458\\
	 mI & 0.922 & 0.996 & 0.00832 & 0.833 & 0.974 & 0.0196\\
	 NAA & 1.00 & 0.997 & 0.00759 & 0.853 & 0.958 & 0.0257\\
	 NAAG & 1.02 & 0.985 & 0.0183 & 0.824 & 0.870 & 0.0472\\ 
	 PCh & 0.857 & 0.944 & 0.0303 & 0.639 & 0.842 & 0.0413\\ 
	 PE & 0.903 & 0.995 & 0.00932 & 0.767 & 0.959 & 0.0229\\  
	 Tau & 0.860 & 0.984 & 0.0159 & 0.629 & 0.887 & 0.0331\\
	 Thr & 1.06 & 0.979 & 0.0222 & 0.782 & 0.819 & 0.0554\\
	 \bottomrule
\end{tabular}
\end{minipage}}
\caption{\small A quantitative comparison between the quantitation results for two different noise levels is shown. A perfect quantitation algorithm would produce the following results: slope = 1, $r^2$ = 1, and standard error (Std. Error) = 0. From the results, it is clear that higher SNR spectra produce more accurate quantitative results. }
\label{table:fittab}
\end{table}

\section*{Discussion}
From the results, it is clear that the D-UNet architecture is capable of both reconstructing non-uniformly sampled L-COSY data and quantifying L-COSY spectra after appropriate training. Figures \ref{fig:NUSrecon} and \ref{fig:quant} show this qualitatively whereas tables \ref{table:MSE} and \ref{table:fittab} show this quantitatively. While deep learning has very recently been used for quantitation of 1D MRS\cite{hatami2018magnetic}, to our knowledge this is the first application of deep learning for reconstructing and quantifying L-COSY MRS. For reconstruction at high acceleration factors, the D-UNet method greatly outperforms a standard compressed sensing method. Spectral quality plays a large role in determining the outcome of the quantitation method, and poor spectral quality results in higher errors, as seen in figure \ref{fig:MSEnoise}. Even though the model architecture for both applications is identical, the two models learn separate properties of the L-COSY spectrum.

The first model, which reconstructs NUS L-COSY data, learns to remove the artifacts produced from the application of a particular non-uniform sampling mask. Due to the non-uniform $t_1$ sampling, various ridging artifacts are present in the $F_1$ domain\cite{schmieder1993application}. Depending on the sampling pattern, the artifacts will be mostly constant for each metabolite, but will still be a function of the metabolite concentration, line-broadening factor, and noise level. By providing enough example data, the network essentially learns how to identify the ridging artifacts and remove them appropriately for a given sampling mask and basis set. This is best illustrated in figure \ref{fig:NUSrecon}, where it is clear that ridging is removed in the reconstructed spectra for each acceleration factor. 

On the other hand, the second model that quantifies metabolite concentrations from L-COSY spectra learns a different property of the L-COSY images. After adding all of the metabolites together to form a composite spectrum, several signals overlap and are hard to disentangle. Optimization problems are able to handle this issue by fitting overlapping peaks using several parameters, often including appropriate prior-knowledge\cite{schulte2006profit,fuchs2014profit,martel2015localized}. Unfortunately, these algorithms take a very long time to calculate these parameters and often yield sub-par results if the quality of the L-COSY spectrum is low (high noise, low SNR, signal contamination, etc.). The quantitation model learns how to disentangle overlapping signals through analysis of the magnitude, real, and imaginary components of the input spectrum. By training on thousands of data, the model learns which signals best represent each metabolite even if the signal is buried in another peak and noise. Furthermore, the calculation is extremely fast and is on the order of seconds for a single spectrum. In terms of accuracy, even lower concentrated metabolites simulated at less than 1 mmol are accurate (error less than 3\%) for most metabolites, as shown in figure \ref{fig:fit}. While these pilot results look promising for reconstruction and quantitation of L-COSY spectra, the current implementation of this method has several weaknesses.

First, the D-UNet model requires prior-knowledge for all metabolites present in the tissue for training as well as how these signals are affected by a particular non-uniform sampling pattern. Compressed sensing reconstructions do not require any spectral prior-knowledge, and therefore are more versatile for different sampling masks. This is not necessarily a weakness if: 1) the sampling mask used for acquisition matches the D-UNet sampling mask used for training and 2) all metabolites in the tissue are known \textit{a priori}. For most experiments, 1) is easily satisfied. For healthy tissues and well documented pathologies, 2) is not an issue. However, 2) may become an issue for pathologies that are not well understood and involve unknown chemical changes. This problem may be alleviated by including prior-knowledge for all metabolites appearing in the analysis of \textit{ex vivo} tissue samples of this pathology if available. For example, mass spectrometry of \textit{ex vivo} tissues played a pivotal role in identifying 2-hydroxyglutarate (2HG) in certain glioma patients as a metabolite of interest\cite{dang2009cancer}. Additional prior-knowledge can always be included into the training process to account for macromolecule signals or other signals that may be present in the spectrum retrospectively if necessary. 

Another weakness of the current methodology is that water and fat contamination were not added to the training spectra. Due to water suppression pulses\cite{ogg1994wet}, spectral distortions around the water region may affect metabolite quantitation. For 2D experiments, total removal of water signal while retaining metabolite signal is more challenging, and may affect the amplitudes of correlated cross-peaks close to water. This problem can be overcome through more advanced training, however the effects of water suppression and removal through common methods such as singular value decomposition (SVD) have to be well understood in order to be modeled correctly. Contaminating fat signal may affect quantitation of metabolites such as lactate and NAA, depending on severity. These fat signals can also be incorporated into the training process, however it is important to utilize the correct fat species.

The final weakness of the current methodology is the broadening model used to produce the training and testing data. Currently, only an exponential line-broadening term was used for this pilot study. While exponential line-broadening may be a great first approximation for peak shapes, gaussian, lorentzian and even voigt lineshapes may be present in the final experimental peaks\cite{fuchs2014profit}. Due to the increased number of parameters introduced with these added lineshapes, the training data size would need to be much larger for adequate training of the model. In addition, the number of features present in the model may need to be increased in order to handle the complexity of the additional broadening parameters.

Even with these weaknesses, the methodology presented here can easily be applied to other 2D MRS experiments and to iterative MRS experiments in general. The J-resolved spectroscopy (JPRESS) experiment is another useful 2D MRS technique\cite{ryner1995localized,kreis1996spatially}, and we have pilot results showing that these models apply for this type of experiment (Supplemental Information). Other 2D experiments include the nuclear overhauser effect spectroscopy (NOESY), total correlation spectroscopy (TOCSY), as well as others. Iterative MRS experiments include diffusion weighted spectroscopy\cite{cao2017vivo,nicolay2001diffusion,ronen2015diffusion}, J-editing spectroscopy, and any multi-TE spectroscopy\cite{hurd2004measurement}. In addition, this methodology could be refined for the application of super-resolution spectroscopy, including covariance spectroscopy\cite{bruschweiler2004covariance,iqbal2017covariance}. However, super-resolution may be unnecessary if accurate quantitative results can already be obtained from low resolution spectra.   

Simulation results are certainly powerful for evaluating the feasibility of potential applications, and this study demonstrates that the D-UNet is capable of reconstructing NUS L-COSY data and quantifying L-COSY spectra. However, these methods need to be further validated \textit{in vitro} and \textit{in vivo}. Also, these methods have to be compared to state of the art techniques for each application. For reconstruction, the D-UNet model should ideally be compared to compressed sensing, maximum entropy\cite{mobli2006spectral,burns2014non}, or other reconstruction methods. It is important to note that while many reconstruction methods require certain sampling schemes (random, non-uniform, etc.), the D-UNet is capable of reconstructing any sampling pattern with the correct training approach. While an exponential sampling scheme was used in this study, a skewed-squared sine-bell sampling scheme may be better to implement in the future\cite{burns2014non}. For quantitation, it is important to compare the deep learning method to other 2D \textit{in vivo} fitting algorithms to assess accuracy and reproducibility\cite{martel2015localized}. After further validation, these models may easily be combined together to create a single deep learning model capable of simultaneously reconstructing and quantifying L-COSY spectra. With further improvements, this method will hopefully have the same acquisition duration as a 1D single-voxel scan (3-5 minutes), which will make this method extremely useful clinically for discerning overlapping metabolite signals.           

\section*{Conclusion}
We present a deep learning approach capable of reconstructing non-uniformly sampled L-COSY spectra and quantifying fully sampled L-COSY spectra. Overall, the results demonstrate accurate reconstruction and quantitation with normalized mean squared error less than 5\% for most SNR levels. This technique was evaluated using simulated data, and further studies will validate this method for \textit{in vitro} and \textit{in vivo} measurements, and compare this method to state of the art techniques.

\newpage

\section*{Supplementary Figures}
The above methodology was also performed for JPRESS experiments. Some example results are shown below, and these results demonstrate the capabilities of the D-UNet for JPRESS acceleration and JPRESS quantitation.
\begin{figure}
	\center
	\hspace*{-0.5cm}
	\includegraphics[scale=0.45]{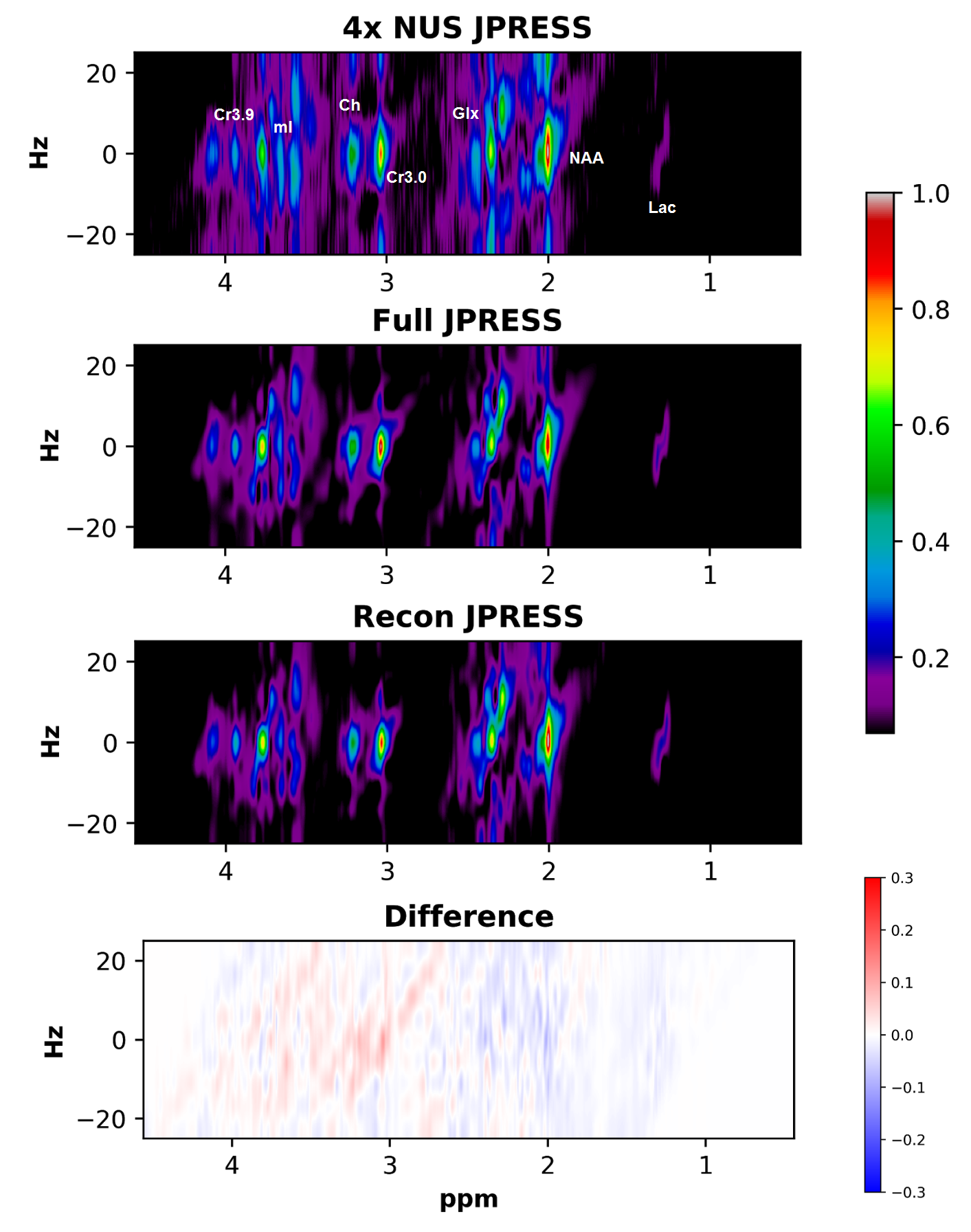}
	\caption{\small A JPRESS spectrum was simulated, and was under-sampled using the same 4x NUS mask shown in figure \ref{fig:NUSrecon}. The actual JPRESS spectrum is shown (Full JPRESS) and is qualitatively compared against the JPRESS spectrum reconstructed using a trained D-UNet model (Recon JPRESS). The difference between the two spectra is also shown (bottom). }
	\label{fig:jpress1}
\end{figure}

\begin{figure}
	\center
	\hspace*{-1.3cm}
	\includegraphics[scale=0.195]{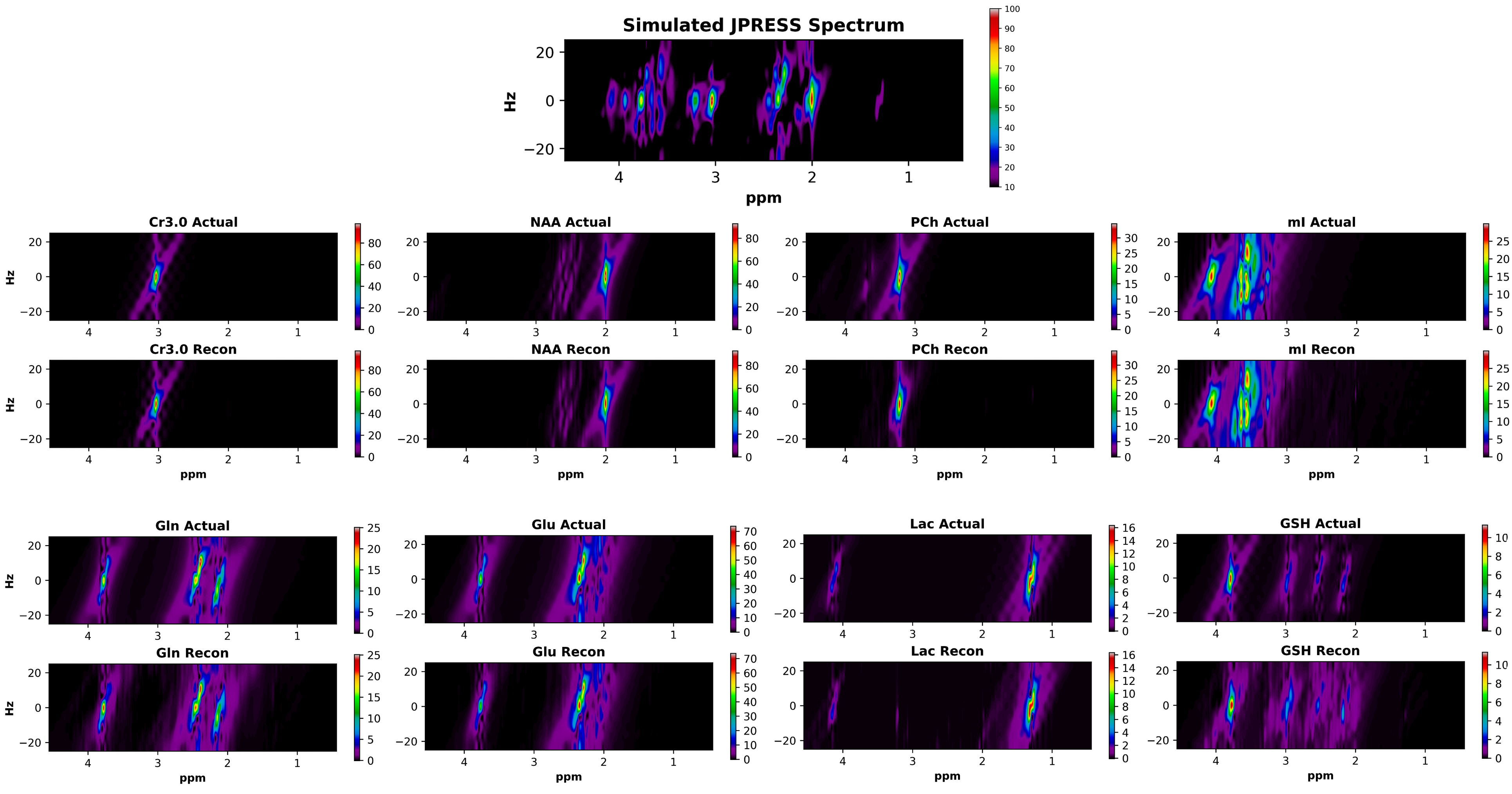}
	\caption{\small A ground truth simulated JPRESS spectrum is displayed. Below the full spectrum are the actual and reconstructed individual metabolite spectra for Cr3.0, NAA, PCh, mI, Gln, Glu, Lac, and GSH. The reconstruction was performed in an identical manner to the L-COSY quantitation described above.}
	\label{fig:jpress2}
\end{figure}

\end{document}